\documentclass[letterpaper, preprint, paper,11pt]{AAS}	

\usepackage{bm}
\usepackage{amsfonts,amsmath,amsthm,amssymb}
\usepackage{subfigure}
\usepackage[colorlinks=true, pdfstartview=FitV, linkcolor=black, citecolor= black, urlcolor= black]{hyperref}
\usepackage{overcite}
\usepackage{footnpag}			      	

\usepackage{mathtools}
\newtheorem{prob}{Problem}

\PaperNumber{XX-XXX}

\begin{document}

\title{Verifiable mission planning for space operations}

\author{%
  Quentin Rommel\thanks{Equal Contribution.} \thanks{Graduate Research Assistant, Department of Aerospace Engineering and Engineering Mechanics, The University of Texas at Austin, 2617 Wichita St, Austin, TX 78712.} ,  
  Michael Hibbard\footnotemark[1] \thanks{Flight Algorithms Engineer, Starfish Space, 665 Andover Park W, Tukwila, WA 98188. (Work conducted as a Graduate Research Assistant, Department of Aerospace Engineering and Engineering Mechanics, The University of Texas at Austin, 2617 Wichita St, Austin, TX 78712.)} ,
  Pavan Shukla\footnotemark[2] ,
  Himanshu Save\thanks{Sr. Research Scientist, Center for Space Research, The University of Texas at Austin, 3925 W. Braker Lane, Suite 200, Austin, TX 78759.} ,
  Srinivas Bettadpur\thanks{Professor, Department of Aerospace Engineering and Engineering Mechanics, The University of Texas at Austin, 2617 Wichita St, Austin, TX 78712.} ,
  and Ufuk Topcu\footnotemark[5]%
}

\maketitle

\begin{abstract}

Spacecraft must operate under environmental and actuator uncertainties while meeting strict safety requirements. Traditional approaches rely on scenario-based heuristics that fail to account for stochastic influences, leading to suboptimal or unsafe plans. We propose a finite-horizon, chance-constrained Markov decision process for mission planning, where states represent mission and vehicle parameters, actions correspond to operational adjustments, and temporal logic specifications encode operational reach–avoid constraints. We synthesize policies that optimize mission objectives while ensuring constraints are met with high probability. Applied to the GRACE-FO mission, the approach accounts for stochastic solar activity and uncertain thrust performance, yielding maneuver schedules that maximize scientific return and provably satisfy safety requirements. We demonstrate how Markov decision processes can be applied to space missions, enabling autonomous operation with formal guarantees.

\end{abstract}

\section{Introduction}
Planning spacecraft operations is critical for sustaining mission objectives, as missions must adapt to variable conditions and limited resources. Current in-operation approaches rely on precomputed scenarios or heuristic scheduling rules that assume fixed environmental conditions\cite{save2023gracefo}. These methods often fail to capture the stochastic nature of orbital decay and space weather variability, leading to inefficient fuel use or violations of operational constraints.

The GRACE-FO mission provides a motivating case study. Designed to map Earth’s gravity field through tandem satellite operations, it has faced two major challenges: malfunctioning thrusters limiting orbit control, and the failure of one satellite’s accelerometer. To mitigate the resulting degradation in measurement accuracy, the mission must operate at higher altitudes, where atmospheric drag is reduced. This requirement, combined with uncertain solar activity and fuel limitations, makes orbit planning particularly difficult. Previous orbit management strategies for GRACE-FO, such as Save et al.~\cite{save2023gracefo} work, rely on fixed solar cycle scenarios and do not capture variability in solar flux, which can result in premature reentry under adverse conditions.

We address these challenges by formulating spacecraft mission planning as a finite-horizon Markov decision process (MDP), a framework for sequential decision making under uncertainty \cite{puterman2014markov}. For GRACE-FO, states capture mission-relevant quantities such as orbital altitude and fuel reserves, actions represent discrete maneuver options, and transition probabilities model the effects of uncertain factors like solar flux and thrust variability. The reward function encodes mission objectives, such as maintaining observation quality at higher altitudes and avoiding repeated groundtrack orbits. Solving the MDP yields maneuver policies that maximize mission return while enforcing constraints such as minimum altitude and maneuver spacing. Unlike traditional scenario-based methods, this approach directly incorporates uncertainty into the planning model, enabling adaptive maneuver scheduling with formal, probabilistic safety guarantees.

Prior studies on mission planning under uncertainty have approached the problem from several directions. Scenario-based analyses, such as the study by Save et al.~\cite{save2023gracefo}, evaluate a fixed set of worst-case, nominal, and best-case conditions, but this often produces overly conservative strategies that lack responsiveness to evolving uncertainties and cannot adapt when conditions deviate from those cases. Another approach is to use stochastic optimization and control techniques~\cite{chai2019stochastic,echigo2025autonomy,ozaki2020tube}. These methods typically operate over short horizons and emphasize local constraint handling, making them less suitable for multi-year, high-level mission planning with operational constraints. At longer horizons, decision-rule synthesis has been used to enable flexible logistics policies under uncertainty~\cite{chen2021flexibility}. However, this approach targets resource allocation rather than closed-loop planning and does not enforce formal safety constraints. In contrast, MDPs provide a sequential decision-making model that naturally supports uncertainty, policy synthesis, and enforcement of discrete safety or timing rules. Prior applications of MDPs include satellite task planning, maneuver scheduling, and resource management under uncertainty~\cite{eddy2020markov,kuhl2025markov,rommel2025optimal,hibbard2025autonomous}. However, these efforts generally encode constraints heuristically or via soft penalties, optimizing expected performance without certifying that safety-critical constraints are satisfied with high probability. Temporal logic has been used for formal specification~\cite{hibbard2023trajectory}, but automaton-based methods require forming a product between the MDP state space and the specification automaton (and, for time-varying rules, an explicit time index), which increases complexity and limits scalability for long-term planning.

We propose a finite-horizon, time-dependent MDP for probabilistic mission planning in spacecraft operations, embedding environmental and system uncertainties. Operational rules are encoded as reach–avoid constraints, and safety is verified via backward dynamic programming on the state–time space, computing satisfaction probabilities without a specification-automaton product~\cite{hibbard2023trajectory,baier2008principles,forejt2011automated}. We apply this approach to GRACE-FO, demonstrating policies that meet mission objectives while satisfying operational safety requirements (e.g., minimum altitude). This extends scenario-based orbit management~\cite{save2023gracefo} and advances decision-theoretic planning for spacecraft operations~\cite{eddy2020markov}, providing formal safety guarantees in space mission planning.




\section{Probabilistic Policy Synthesis and Safety Verification Using Markov Decision Processes}

We model spacecraft decision-making under uncertainty as a finite-horizon Markov decision process (MDP). An MDP is a common paradigm for modeling planning and acting in stochastic environments with nondeterministic action selection~\cite{puterman2014markov}.
For mission planning, we focus on \textit{finite-horizon} MDPs, wherein the decision-making process is performed over a finite number of time steps called a \textit{planning horizon}.
In the context of a spacecraft mission, the environment stochasticity could be due to, e.g., uncertainties in future space weather conditions, while actions could correspond to performing altitude raises.
Similarly, the planning horizon could be the operational lifetime of the mission, while individual time steps are specific points in time at which actions are chosen, e.g., daily, weekly, or monthly.

Formally, a finite-horizon MDP $\mathfrak{M}$ is described by the tuple $\mathfrak{M} = \langle \mathcal{S} , \mathcal{A} , \mathcal{H} , T , R , \mathbf{P}_{0} \rangle$,
where $s \in \mathcal{S}$ is a finite set of states, $a \in \mathcal{A}$ is a finite set of actions, $\mathcal{H} = \{0,1,2,\ldots,h,\ldots,H \}$ is a set of time indices with planning horizon $H < \infty$, $T:\mathcal{S} \times \mathcal{A} \times \mathcal{H} \times \mathcal{S} \rightarrow [0,1] $, 
is a transition function mapping state-action pairs at time step $h$ to a probability distribution over successor states, $R: \mathcal{S} \times \mathcal{A} \times \mathcal{H} \rightarrow \mathbb{R}$ is a reward function mapping state-action pairs at time step $h$ to an associated reward, and $\mathbf{P}_{0}:\mathcal{S}\rightarrow [0,1]$ is an initial state distribution. 
Let $R(s,a,h)$ denote the reward obtained for selecting action $a$ while in state $s$ at time step $h$.
Similarly, let $T(s'|a,s,h)$ denote the probability of transitioning to state $s'$ after taking action $a$ while in state $s$ at time step $h$.
Note that since the transition function must define a valid probability distribution, we have $\sum_{s' \in \mathcal{S}} T(s'|a,s,h) = 1$ for all $(s,a,h) \in \mathcal{S} \times \mathcal{A} \times \mathcal{H}$.

\begin{figure*}
    \centering
    \includegraphics[width=0.8\columnwidth]{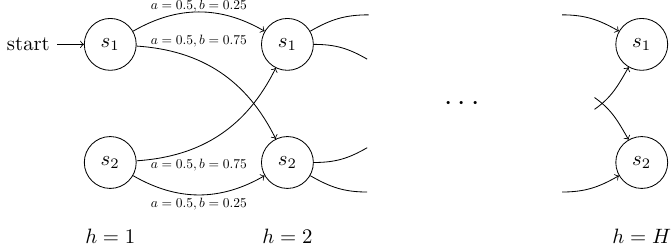}
    \caption{A visualization of a simple finite-horizon MDP with horizon length $H$. The set of states is $\mathcal{S}=\{ s_{1} , s_{2} \}$} and the set of actions is $\mathcal{A} = \{ a , b\}$.
    \label{fig:example_mdp}
\end{figure*}
A visualization of a simple finite-horizon MDP is shown in Figure~\ref{fig:example_mdp}, where the planning horizon is of length $H$.
The set of states is $\mathcal{S} = \{ s_{1} , s_{2} \}$, the set of actions is $\mathcal{A} = \{ a, b \}$, and the initial state is $s_{1}$.
For the initial time step, each state-action pair is labeled by its associated transition probability, e.g., $T(s_{2}|s_{1},b,1) = 0.75$.
Intuitively, the transitions in the MDP flow only forward in time.

To resolve the nondeterminism in choosing an action, the decision-maker must synthesize a \textit{time-dependent policy} $\pi \in \Pi$ \, where $\Pi$ is the set of time-dependent policies.
Formally, $\pi$ is a sequence of decision rules $\{d_{1},d_{2},\ldots , d_{H}\}$, where each $d_{h}:\mathcal{S} \times \mathcal{A} \rightarrow [0,1]$ prescribes a probability distribution for action selection in each state at time step $h$.  A policy is simply a rule that can specify, for example, which maneuver a spacecraft should execute given its current state and time.
Notice that each decision rule is a function of only the current state of the decision-maker, rather than of its entire history of states.
Such a policy is referred to as \textit{Markovian}.
Let $\pi(a|s,h)$ denote the probability that action $a$ is selected while in state $s$ at time step $h$ under policy $\pi$
%
If there exists an action $a \in \mathcal{A}$ for each $(s,h) \in \mathcal{S} \times \mathcal{H}$ such that $\pi(a|s,h) = 1$ and $\pi(a'|s,h) = 0$ for all $a' \in \mathcal{A} \setminus a$, we refer to $\pi$ as \textit{deterministic}.
%
%
Finally, for a time-dependent policy, let $\pi(s,h)$ denote the action selected deterministically while in state $s$ at time step $h$.

Given the spacecraft mission environment expressed in terms of an MDP, we seek to determine whether there exists a policy that satisfies a given mission specification with probability at least $1 - \delta$. 
%
%
We want the spacecraft to always remain safe. In other words, it should never enter an unsafe set, denoted as $\mathcal{B} \subseteq \mathcal{S}$. We require that the policy guarantees safety with a probability of at least $1-\delta$. Here, $\delta \in [0,1)$ represents the maximum allowed chance of a safety violation.
The safety probability requirement for a policy $\pi \in \Pi$ associated to the specification $\varphi$ can be written as:
\begin{align}
\mathrm{Pr}^{\pi}(s_0 \models \square \neg \mathcal{B}) &\geq 1-\delta,
\end{align}
where $\square \neg \mathcal{B}$ means “Always avoid $\mathcal{B}$”, and $s_0 \in \mathcal{S}$ is the initial state of the MDP. Equivalently, since “always avoid $\mathcal{B}$” is the opposite of “eventually reaching $\mathcal{B}$”, we can also write an equivalent reachability specification as:
\begin{align}
\mathrm{Pr}^{\pi}(s_0 \models \lozenge \mathcal{B}) &\leq \delta.
\end{align}

We assume that the mission of the spacecraft is to maximize the total expected reward collected over the planning horizon subject to a constraint that a mission safety specification $\varphi$ must be satisfied with probability at least $1 - \delta$.
%
%
We now formally state the problem of the decision-maker.
\begin{prob}\label{prob:main_prob} Given an MDP $\mathfrak{M}$ representing the spacecraft environment and a mission safety specification $\varphi$, synthesize a policy $\pi^{*}$ that solves the following optimization problem.
\begin{subequations}
\begin{align}
    \max_{\pi \in \Pi} & \qquad \mathbb{E} \left[ \sum\nolimits_{h=1}^{H} R(s,a,h )\right] \label{eq:main_prob_1} \\
    \mathrm{s.t.} & \qquad \mathrm{Pr}^{\pi}( s_0 \models \varphi) \geq 1 - \delta,\label{eq:main_prob_2}
\end{align}
\end{subequations}
where $\mathrm{Pr}^{\pi}( s_0 \models \varphi)$ is the probability that the mission safety specification $\varphi$ is satisfied under policy $\pi$ starting in state $s_0$ in MDP $\mathfrak{M}$.
\end{prob}
This problem is solved in two steps. First, the policy $\pi\in\Pi$ for the objective~\eqref{eq:main_prob_1} is synthesized without considering the constraint~\eqref{eq:main_prob_2}.
Formally,
\begin{align}\label{eq:optimal_policy}
    \pi^{*} = \mathrm{arg}\max_{\pi \in \Pi} \mathbb{E}\left[ \sum\nolimits_{h=1}^{H} R(s,a,h) \right],
\end{align}

is obtained by solving the recursive Bellman equations backwards-in-time according to
\begin{subequations}
\begin{align}
    V^{*}(s,H) & = \max_{a\in\mathcal{A}} \left[ R(s,a,H) \right]      \quad  & \forall s \in \mathcal{S}, \label{eq:recursive_bellman_1} \\
    V^{*}(s,h) & = \max_{a\in\mathcal{A}} \left[ R(s,a,h) + \sum\nolimits_{s' \in S} T(s'|s,a,h) V^{*}(s',h+1) \right] \quad & \forall (s,h) \in  \mathcal{S} \times \mathcal{H} \setminus H, \label{eq:recursive_bellman_2} 
\end{align}
\end{subequations}
where $V^{*}(s,h)$ is the optimal \textit{value function} for state $s$ at time step $h$ and $\pi^{*}(s,h)$ is the corresponding maximizing value of $a \in \mathcal{A}$.
We then evaluate the safety performance of the obtained policy $\pi$. First, we define
$
V^\varphi(s,h)= \mathrm{Pr}^{\pi}((s,h) \models \square \neg \mathcal{B}).
$
Note that here $\pi$ is the policy obtained from solving the unconstrained problem~\eqref{eq:optimal_policy}. The safety value function $V^\varphi(s,h)$ is computed using the following formulation:
\begin{align}
   & V^\varphi(s,h) = 
    \begin{cases}
    0, & \text{if } s \in B, \\
    1, & \text{if } s \not\models \exists \lozenge \mathcal{B}, \\
    \displaystyle \max_{a \in \mathcal{A}} \sum_{s'\in \mathcal{S}} T(s'|s,a,h) \, V^\varphi(s',h+1), & \text{otherwise,}
    \end{cases}
    \label{eq:specification_value_function}
\end{align}
with an appropriate terminal condition at $h=H$ (i.e. for $s \notin \mathcal{B}$, $V^\varphi(s,H) = 1$ if no further transitions occur). A policy is considered feasible if it guarantees that the safety probability at the initial state is at least $1-\delta$:
\begin{align}
    V^\varphi(s_0,0) &\geq 1-\delta,
\end{align}
ensuring that the spacecraft remains safe with the required probability starting from $s_0$.

To model the safety specification $\varphi$, we can characterize it in terms of a \textit{temporal logic formula}~\cite{baier2008principles, rozier2011linear}.
The foundational building blocks of a temporal logic formula are \textit{atomic propositions}, which evaluate to either \texttt{true} or \texttt{false}, e.g., ``\textit{the spacecraft has nonzero fuel reserves remaining}.''

Note that it is straightforward to generalize the constraint~\eqref{eq:main_prob_2} to the probability of multiple mission specifications. We refer the reader to Forejt et al.\cite{forejt2011automated} section 8 for further details.

\section{Space Environment and Dynamics}
Transitioning from the abstract MDP formulation to a physically accurate representation of the operational environment is essential for synthesizing meaningful policies. While the MDP provides a framework for decision-making under uncertainty, its fidelity depends on a precise characterization of the forces governing spacecraft motion. Orbital mechanics forms the foundation for modeling a spacecraft’s trajectory by accounting for gravitational forces, atmospheric drag, and other perturbations, which in turn define the deterministic states, actions, and transitions in our MDP. In contrast, space weather introduces stochastic effects: its variability is difficult to predict yet has a direct impact on the operational environment. These uncertainties are incorporated into the transition probabilities, ensuring that the resulting policies remain robust to both predictable dynamics and unpredictable environmental fluctuations.

\subsection{Orbital Mechanics}

First, we introduce key ideas from orbital mechanics and then present how we can build an MDP for space mission planning.
Consider a spacecraft in an elliptical orbit around the Earth.
Let $\mathbf{r}(t) = [\mathbf{p}(t)^{\top}, \mathbf{v}(t)^{\top}]^{\top} \in \mathbb{R}^{6}$ denote the Cartesian state of a spacecraft in the inertial frame, where $\mathbf{p}(t) = [x(t),y(t),z(t)]^{\top}$ and $\mathbf{v}(t) = [\dot{x}(t),\dot{y}(t),\dot{z}(t)]^{\top}$ are the spacecraft position and velocity at time $t$.
%
In what follows, we will omit the time index $t$ for notational clarity.
Assuming a spherical Earth, the altitude $R_{\mathrm{alt}}$ of the spacecraft above Earth's surface can subsequently be obtained from its position $\mathbf{p}$ according to
\begin{align}\label{eq:altitude}
    R_{\mathrm{alt}} = ||\mathbf{p}||_{2} - R_{e},
\end{align}
where $R_{e}$ is the radius of the Earth. 
The dynamics of the spacecraft are given by
\begin{align}\label{eq:eom_with_drag}
    \ddot{\mathbf{r}} = -\mu \frac{\mathbf{r}}{||\mathbf{r}||_{2}^{3}} + F_{\mathrm{drag}} + F_{\mathrm{pert}},
\end{align}
where $\mu$ 
is the gravitational parameter of the Earth, $F_{\mathrm{drag}}$ is the atmospheric drag force, and $F_{\mathrm{pert}}$ is the combination of the remaining perturbing forces acting on the spacecraft.
The term $F_{\mathrm{pert}}$ could include, for example, $J2$ and $J3$ effects.
The atmospheric drag term $F_{\mathrm{drag}}$ is equal to
\begin{align}\label{eq:atmospheric_drag}
    F_{drag} \triangleq -\frac{1}{2} C_{d} \frac{A}{m} \rho_{A} v_{a} \mathbf{\textit{v}}_{a},
\end{align}
where $C_{D}$ is the drag coefficient of the spacecraft, $A$ is the cross-sectional area of the spacecraft,
$m$ is the mass of the spacecraft,
$\rho_{A}$ is the altitude-dependent atmospheric density,
%
%
$\mathbf{\textit{v}}_{a}$ is the velocity of the spacecraft relative to the Earth-fixed frame, given by
\begin{align*}
    \mathbb{\textit{v}}_{a} \triangleq \begin{bmatrix}
        \dot{x} + \dot{\theta} y \\
        \dot{y} - \dot{\theta} x \\
        \dot{z}
    \end{bmatrix}, \quad v_{a} = || \mathbb{\textit{v}}_{a}||_{2},
\end{align*}
in which $\dot{\theta}$
is the rotation rate of the Earth.
%
The atmospheric density $\rho_{A}$ is evaluated using the NRLMSISE-00 model~\cite{picone2002nrlmsise,emmert_nrlmsis_2021,emmert_nrlmsis_2022} as a function of altitude, and the solar flux at 10.7 cm (F10.7) obtained from NOAA predictions~\cite{miesch2024solar}.

%

\begin{figure}
    \centering
    \includegraphics[width=0.5\columnwidth]{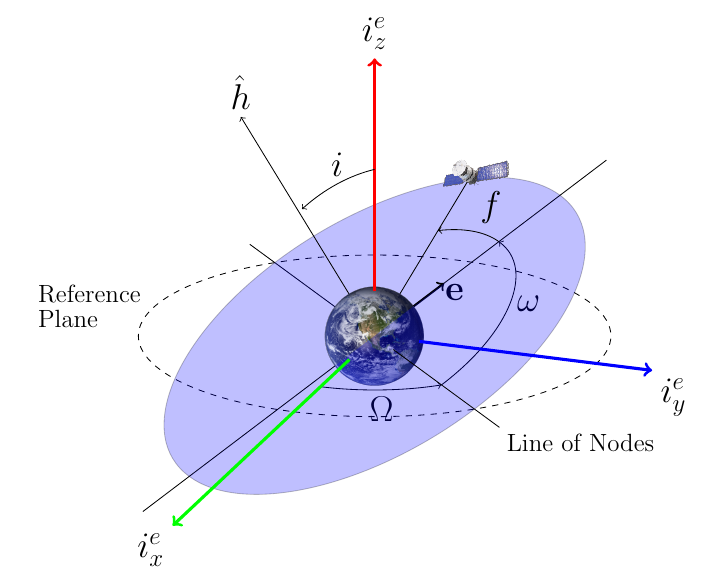}
    \caption{Visualization of the Keplerian orbital elements.}
    \label{fig:orbital_elements}
\end{figure}

In tandem with the preceding Cartesian representation of the spacecraft dynamics, one can instead characterize its orbit in terms of the Keplerian orbital elements $\{a,e,i,\omega,\Omega,f\}$, where $a$ is the semimajor axis length, $e \in [0,1)$ is the eccentricity, $i$ is the inclination, $\omega$ is the argument of periapse, $\Omega$ is the longitude of the ascending node, and $f$ is the true anomaly~\cite{prussing1993orbital}.
%
%
These orbital elements are visualized in Figure~\ref{fig:orbital_elements}, where the angles are defined with respect to the Earth-centered inertial reference frame.
We refer the reader to~\cite{prussing1993orbital} for information on how to convert between the Cartesian and Keplerian representations of the spacecraft state.

\subsection{Space Weather}

Due to the atmospheric drag force $ F_{\mathrm{drag}}$, the eccentricity and semimajor axis of the spacecraft orbit decrease monotonically over its operational lifetime, eventually approaching values of $R_{e}$ (Earth’s radius) and zero, respectively. As a result, the spacecraft gradually descends to a circular orbit of zero altitude~\cite{prussing1993orbital}. The extent and variability of this degradation depend heavily on the atmospheric density at the spacecraft's altitude, which is strongly influenced by solar activity.

The solar cycle governs the Sun’s activity levels and, in turn, the atmospheric environment encountered by the spacecraft. However, the duration and intensity of individual cycles vary; larger cycles tend to last longer than 11 years, while smaller cycles are typically shorter. Solar activity is commonly categorized into three levels: high, moderate, and low \cite{save2023gracefo}. During high solar activity, increased ultraviolet and X-ray radiation heat and expand the Earth's upper atmosphere, raising atmospheric density at orbital altitudes and increasing drag forces. This accelerates orbital decay, requiring more frequent station-keeping maneuvers and increasing fuel consumption. In moderate activity periods, atmospheric density variations are less extreme, allowing for more stable orbit maintenance. Conversely, low solar activity leads to reduced drag, minimizing fuel expenditure and extending mission lifetime.


The variability in solar activity across different cycles can be incorporated into the MDP transition model by defining states and transitions that reflect the probabilistic nature of future solar activity levels. Since larger solar cycles exhibit prolonged high-drag conditions, while smaller cycles may allow for extended low-drag periods, transitions between states can be weighted using historical solar cycle data. By integrating probabilistic dependencies into the MDP, the model allows for dynamic adjustments in altitude control, optimizing fuel usage, and ensuring prolonged mission operation under different space weather conditions.

%




\section{Case study}

\begin{figure}
    \centering
    \includegraphics[width=0.4\columnwidth]{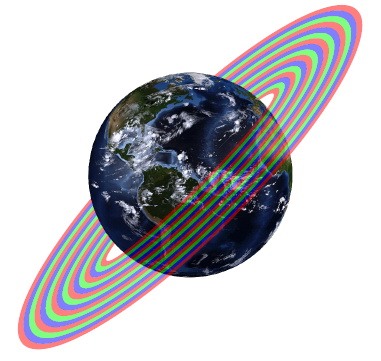}
    \caption{We construct the discrete representation of the mission environment by sampling a representative set of mean orbits.}
    \label{fig:disc_semimajor}
\end{figure}

Using the MDP theory and the described dynamics, we now present how to construct the MDP model that will be used in solving Problem~\ref{prob:main_prob} in the context of the GRACE-FO mission.

\subsection{GRACE-FO mission}
The Gravity Recovery and Climate Experiment (GRACE), launched in 2002, was a NASA mission that sought to obtain high-accuracy estimates of Earth's gravity field from twin satellites in near-polar, near-circular orbits~\cite{tapley2004gravity}. 
The GRACE-FO (Gravity Recovery and Climate Experiment Follow-On) mission, launched in May 2018, continues the legacy of the original GRACE mission to measure Earth's gravity field with unprecedented precision. By monitoring variations in the inter-satellite range using K-Band Ranging and Laser Ranging Interferometer technologies, GRACE-FO provides critical data on global surface mass changes. 

One major challenge arises from the failure of the accelerometer on GRACE-FO 2 (GF2), rendering it unable to measure non-gravitational forces directly. To compensate, the data from the accelerometer of GRACE-FO 1 (GF1) are transplanted to GF2, relying on the assumption that both satellites experience nearly identical environmental conditions within the $
\sim 24–30$ s separation delay. The transplant method recovers differential non-gravitational forces with high fidelity\cite{harvey2024graced} under nominal conditions, but it introduces signal-proportional errors that can reach a few percent of the drag signal in high-drag regimes, which degrades the resulting gravity fields. 


Propulsive capability is another constraint. Altitude-raising was not a design objective, so the cold-gas system and fuel budget are limited for orbit-raising. In addition, thrust realization during burns is uncertain (mismatch between commanded and actual impulse), which introduces variance in achieved altitude change and propellant usage. With finite fuel, this creates a trade-off between performing raises to preserve data quality and conserving propellant for mission lifetime.

Finally, solar-cycle variability drives large, stochastic changes in upper-atmospheric density and hence drag. Periods of elevated activity accelerate orbital decay and amplify transplant-related errors; quieter periods relax both.

We seek a planning strategy that addresses three coupled challenges: the trade-off between altitude and measurement fidelity, limited fuel with uncertain thrust performance, and stochastic drag driven by solar activity. To this end, we present a finite-horizon MDP formulation with altitude and fuel as state variables, orbit-raising maneuvers as discrete actions, and transition dynamics derived from drag and thrust-realization models. The resulting formulation enables optimization of science return while ensuring compliance with safety constraints.

\subsection{State Modeling}


We discretize the environment of the spacecraft in a pair of dimensions.
The first of these dimensions is the altitude of the spacecraft.
Recall that we are primarily interested in the effect of solar flux on atmospheric drag, and that atmospheric drag only affects the semimajor axis length and the eccentricity of the spacecraft orbit.
Furthermore, since the GRACE-FO mission is nominally in a circular orbit (i.e. $e = 0$), its eccentricity is unaffected by atmospheric drag and remains constant over time.
Thus, we focus solely on discretizing the altitude of the spacecraft (which is a function of the semimajor axis length as given by~\eqref{eq:altitude}).

To construct this discretization, let $a_{min}$, $a_{max} \in \mathbb{R} \geq 0$, $a_{max} > a_{min}$, be a minimum and maximum altitude, respectively, and $N_{a} \in \mathbb{Z}$ denote the number of discrete altitudes to consider within these bounds.
Denote the set of these discrete altitudes by $\mathcal{S}_a \triangleq \{ a_{min}+\frac{R_{a}}{2} , a_{min} + \frac{3R_{a}}{2} , \ldots, a_{max}-\frac{R_{a}}{2} \}$, where $R_{a} \triangleq \frac{a_{max} - a_{min}}{N_{A}}$.
Each altitude $s \in \mathcal{S}_a$ corresponds to a discrete band centered around $s$, i.e., the band $[s-\frac{R_{a}}{2} , s + \frac{R_{a}}{2}]$, as shown in Figure~\ref{fig:disc_semimajor}.
Although the true semimajor axis length of the spacecraft will not generally reside on one of the sampled altitudes $s \in \mathcal{S}_a$, we can instead bin it into a corresponding band (assuming that it lies within the minimum and maximum altitudes).

The second discretization is in the temporal dimension.
Particularly, given a mission planning horizon $H \in \mathbb{R}$ (nominally expressed in units of seconds) and a number of discrete time steps $N_{H} \in \mathbb{Z}$, let $\mathcal{H} \triangleq \{0,\frac{H}{N_{H}}, \frac{2 H}{N_{H}},\ldots, \frac{(N_{H}-1)H}{N_{H}} \}$ denote the set of discrete time indices.
In what follows, it is assumed that decisions can only be made at time instants belonging to the set $\mathcal{H}$.

To construct the fuel discretization, we use a similar method to that for the altitude. Let $f_{min}, f_{max} \in \mathbb{R}_{\geq 0}$, with $f_{max} > f_{min}$, be the minimum and maximum fuel levels, respectively, and let $N_f \in \mathbb{Z}$ denote the number of discrete fuel levels to consider within these bounds. We denote the set of these discrete fuel levels by  $\mathcal{F} \triangleq \{ f_{min} + \frac{R_f}{2},\; f_{min} + \frac{3R_f}{2},\; \ldots,\; f_{max} - \frac{R_f}{2} \},$ where $R_f \triangleq \frac{f_{max} - f_{min}}{N_f}.$ Following the same procedure as stated for the altitude, the true fuel level of the spacecraft is binned into the corresponding fuel band.

\subsection{Reward Modeling}

At a high level, the objective of the GRACE-FO mission is to provide high-quality observations of changes in Earth's gravitational field~\cite{landerer2020extending}.
The quality of these observations generally increases with lower atmospheric density, which occurs at higher altitudes.
Due to this fact, the reward function $R$ should be independent of the choice of the action selection and only a function of the current state (i.e., altitude) and time step of the mission.
The dependence of the reward function $R$ on time could be due to, e.g., time-varying atmospheric conditions that affect the observation quality. 
Thus, we introduce $R(s,:,h) \triangleq R(s,a,h)$ for all $a \in \mathcal{A}$ to denote the reward (a function of the observation quality) for being in a state $s \in \mathcal{S}$ corresponding to a fixed, discrete altitude. In other words, $R(s,:,h)$ is independent of the action taken.
Note that $R(s,:,h)$ should be non-negative and monotonically increasing as the corresponding altitude of $s$ increases.

\begin{figure} 
   \centering
   \includegraphics[width=0.6\linewidth]{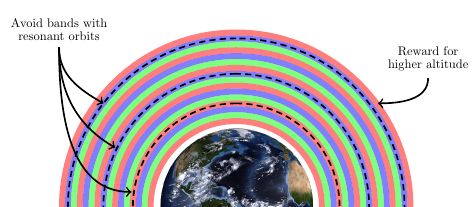}
   \caption{Visualization for the shaping of the reward function $R$.}
   \label{fig:rewards}
\end{figure}

An additional hindrance to observation quality is the existence of \textit{resonant orbits}, which are orbits with a periodic ground track~\cite{pie2008subcycle}.
Due to the periodicity of this ground track, less of Earth's surface is observed, reducing the quality of the global gravitational field estimation.
Thus, for any discrete altitude state $s \in \mathcal{S}_a$ for which a resonant orbit exists in its corresponding altitude band $[s-\frac{R_{a}}{2}, s+\frac{R_{a}}{2}]$, we can set $R(s,:,h) = 0$.
Alternatively, given the periodicity of the resonant orbit $i$, $res_i$ (i.e., the rate at which the ground track repeats), one can scale the reward function by a multiplicative factor $\alpha < 1$ to account for variations in the reduction of observation quality. The reward function can then be defined as
\begin{align}
    R(s,:, h) = \mathbb{E} \left[ alt_{h} \left( \Pi_{i=1} ^ N \left( \alpha \frac{res_i}{res_{max}} + \beta\right)\right)^\frac{1}{N} \right], \label{eq:reward}
\end{align}
Here, $alt_{h+1}$ denotes the spacecraft's altitude at the next time step, $N$ is the number of resonant orbits encountered until that step, and $res_i$ represents the repetition rate of the $i$th resonant orbit, normalized by $res_{max}$, the maximum possible repetition rate. $\alpha$ scales the normalized repetition rate to penalize high resonance, while $\beta$ provides an offset to ensure that the penalty does not reduce the reward excessively. Taking the geometric mean (by raising the product to the power of $\frac{1}{N}$) aggregates the multiplicative penalties, and the expectation $\mathbb{E}[\cdot]$ accounts for uncertainty in the next altitude and resonance effects, thus balancing the benefits of increased altitude with the degradation in observation quality due to repetitive ground tracks.
A visualization of the process for setting the reward for being in a discrete altitude state $s \in \mathcal{S}_a$ is shown in Figure~\ref{fig:rewards}.

Resonant orbits are computed by enforcing the repeat ground track condition 
$$
D \cdot T_{\text{sat}} = N \cdot T_{\text{nodal}},
$$
where $D$ and $N$ denote the number of satellite revolutions and nodal days, respectively. Here, the satellite's nodal period is given by 
$$
T_{\text{sat}} = \frac{2\pi}{n + \dot{\omega}},
$$
where $n$ is the mean motion and $\dot{\omega}$ is the perigee precession rate. The Earth's nodal period is expressed as 
$$
T_{\text{nodal}} = \frac{2\pi}{\omega_e - \dot{\Omega}}.
$$
where $\omega_e$ is the Earth's rotation rate and $\dot{\Omega}$ is the nodal precession rate.
Substituting these into the resonance condition yields 
$$
\frac{D}{n + \dot{\omega}} = \frac{N}{\omega_e - \dot{\Omega}}.
$$
Since the mean motion $n$ is directly related to the semi-major axis via Kepler's law, this equation is solved iteratively using the Lagrange Planetary Equations with Klokočník’s method \cite{klokovcnik2003fine}. We then obtain the resonant altitudes. The resulting normalized repetition rate is then incorporated into the reward function~\eqref{eq:reward}.

\subsection{Action and Fuel Cost Modeling}

\begin{figure}
    \centering
    \includegraphics[width=0.3\linewidth]{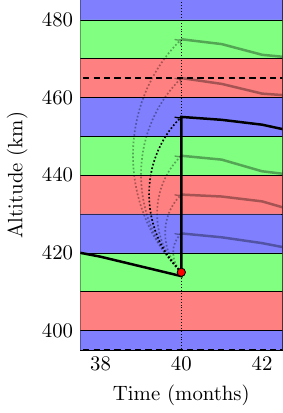}
   \caption{Visualization of the set of feasible actions $\mathcal{A}$, which correspond to different instantaneous altitude gains.}
    \label{fig:actions}
\end{figure}

There are several choices to model the discrete set of actions $\mathcal{A}$ for the GRACE-FO mission.
At their core, these actions represent the choice of how much altitude to gain over the subsequent discrete time interval.
For example, if the temporal discretization is fine enough, then a single, continuous burn could be approximated by a sequence of smaller burns performed at each discrete time step.
In this case, the set of actions is $\mathcal{A} = \{no \, burn , {}^{+}$$R_{a} \}$, where ${}^{+}$$R_{a}$ denotes the choice to increase the altitude by $R_{a}$.
Since the bands are assumed to be evenly spaced in altitude, this action corresponds to transitioning to the state $s' \in \mathcal{S}$ such that the altitude of $s'$ is equal to the altitude of $s$ plus $R_{a}$.
For models with lower temporal resolution, multiple actions corresponding to a finite set of altitude increases can be included in the set $\mathcal{A}$, i.e., $\mathcal{A} = \{ no \, burn , {}^{+}$$R_{a} , {}^{+}$$2R_{a}, \ldots , {}^{+}$$N_{A}R_{a} \}$, where $N_{A}R_{a}$, $N_{A} \in \mathbb{Z}$, is the maximum one-step orbit increase.
A visualization of such a set $\mathcal{A}$ is shown in Figure~\ref{fig:actions}, where there exists a set of seven possible actions (no burn and six possible altitude increases).
The selected action is in black.
A third option for the choice of action is to use a geometric set to approximate the choice of increasingly large fuel burns.
Specifically, let $\hat{\gamma} > 1 $ be a constant scaling factor and $N_{A}$ be the number of actions. 
Then, $\mathcal{A}$ corresponds to the altitude increase $\{ no \, burn, {}^{+} \lceil \hat{\gamma}^{0} \rceil R_{a} , {}^{+} \lceil \hat{\gamma}^{1} \rceil R_{a} , \ldots, {}^{+} \lceil \hat{\gamma}^{N_{A}-1} \rceil R_{a} \}$.

Associated with each action is a fuel cost, while the decision-maker must ensure that the cumulative fuel cost throughout the mission does not exceed its fuel budget.
Let us denote the fuel cost when taking action $a \in \mathcal{A}$ while in a state $s \in \mathcal{S}$ by $C(s,a)$ and the fuel budget by $C_{\mathrm{max}}$.
Trivially, for the action corresponding to \textit{no burn}, the fuel cost is zero.
For the remaining actions, we present two options for modeling the fuel costs and the fuel budget:
\begin{enumerate}
    \item \textit{Maximum Altitude Change.} As previously discussed, each action is associated with an increase in the operational altitude of the GRACE-FO mission. The cost of an action is its corresponding altitude increase, i.e., $C(s,{}^{+}$$nR_{a}) = {}^{+}$$nR_{a}$ for all $n \in \{1,\ldots,N_{A}\}$. Similarly, the fuel budget is a maximum cumulative altitude increase over the mission horizon.
    \item \textit{Fuel usage, $\Delta v$.} Associated with each action in each state is a corresponding fuel usage, encoded by the $\Delta v$ required to complete the altitude change. The fuel budget is a maximum cumulative $\Delta v$ usage over the mission horizon. Due to changes in atmospheric density over time (such as due to varying solar flux), the fuel cost is likewise time-varying, which we express by $C(s,{}^{+}$$nR_{a},h)$.
\end{enumerate}
%
\subsection{Transition Modeling}

Several layers of stochasticity exist in the mission environment.
The first of these layers is the future uncertainty in the solar flux level, i.e., whether the next solar cycle will result in low, medium, or high solar flux.
Denote each of these probabilities by $\mathrm{Pr}_{low}$, $\mathrm{Pr}_{med}$, and $\mathrm{Pr}_{high}$, respectively.

\begin{figure}
   \centering
   \includegraphics[width=0.6\linewidth]{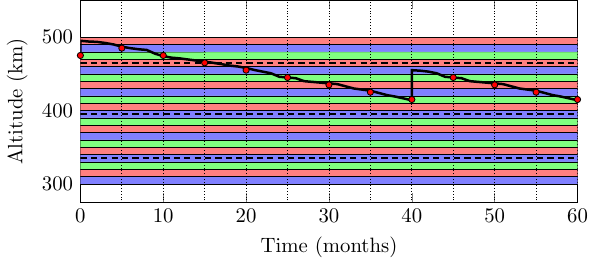}
   \caption{Example visualization of the evolution of the mission altitude over the planning horizon.}
   \label{fig:time_evolution}
\end{figure}

The second layer of uncertainty is that the solar flux over time for a fixed solar cycle intensity level is also stochastic.
Denote the probability of a given value of the solar flux $f$ being realized given a fixed solar cycle intensity by $\mathrm{Pr}(f|i)$, where $i \in \{ low, \, medium, \, high \}$. 
As previously discussed, the solar flux affects atmospheric density, which in turn affects the drag coefficient $C_{D}$.
The larger the coefficient of drag, the more quickly the altitude decays.
Thus, the rate of altitude decay is a random variable that is dependent on the current solar flux.
Assuming an instantaneous altitude change, for a given state-action pair, we can then populate the transition probabilities according to
\begin{subequations}
\begin{align}
    & T(s'|s,a,h) =\\
    & \mathrm{Pr}( \mathrm{alt}_{h+1} \in [R_{s'} -\frac{R_{a}}{2}, R_{s'} + \frac{R_{a}}{2} ] | \mathrm{alt}_{h} = R_{s} + {}^{+}R_{a} ) = \label{eq:tf_1} \\
    &\sum_{i \in \left\{low, medium, high\right\} } \mathrm{Pr}_{i} \int_{f} \mathrm{Pr}(f|i) \cdot \nonumber \\
    &  \mathrm{Pr}( \mathrm{alt}_{h+1} \in [R_{s'} -\frac{R_{a}}{2}, R_{s'} + \frac{R_{a}}{2} ] | \mathrm{alt}_{h} = R_{s} + {}^{+}R_{a} , f) df, \label{eq:tf_2}
\end{align}
\end{subequations}
where $R_{s}$ is the altitude corresponding to state $s \in \mathcal{S}$ and ${}^{+}R_{a}$ is the altitude gain corresponding to selecting action $a \in \mathcal{A}$. 
Note that~\eqref{eq:tf_2} is obtained from~\eqref{eq:tf_1} by explicitly including the conditioning of the transition probability on the solar flux.
Furthermore, since the state dynamics themselves are assumed to have no stochastic disturbances, $\mathrm{Pr}( \mathrm{alt}_{t+1} \in [R_{s'} -\frac{R_{a}}{2}, R_{s'} + \frac{R_{a}}{2} ] | \mathrm{alt}_{t} = R_{s} + {}^{+}R_{a} , f)$ takes either a value of $0$ or $1$ (i.e., it is binary).
In practice, the integral over $f$ can be numerically approximated, and the state transition matrix can be used to efficiently compute the subsequent altitude. A visualization of an example trajectory (with respect to the mission altitude) is shown in Figure~\ref{fig:time_evolution}.

Note that decisions are only made at the discrete time instances represented by the densely dotted vertical lines.
%

In general, a single continuous burn using all remaining fuel is the optimal action to maximize the operational lifetime of the mission.
However, such an action lacks any robustness as no fuel remains aboard the spacecraft to address any potential changes to the mission plan.
Several possibilities to increase the strategy robustness are listed as follows:
\begin{enumerate}
    \item[\textbf{(i)}] \textbf{Collision avoidance.} When the risk of collision with another spacecraft or orbital debris increases, GRACE-FO cannot execute effective avoidance maneuvers. This risk is quantified by a collision rate, $\lambda_c$, whose integration over discrete time intervals yields the probability that a collision avoidance maneuver is required.
    \item[\textbf{(ii)}] \textbf{Station keeping.} In order to maintain the desired orbit, the spacecraft must conserve fuel against perturbing forces that affect not only altitude but also other orbital parameters such as inclination. This requirement is managed by imposing a maximum fuel budget constraint.
    \item[\textbf{(iii)}] \textbf{Fuel leaks.} A critical failure in the fuel system may render the remaining fuel unusable. The likelihood of such a fuel leak is expressed by a fuel failure rate, $\lambda_f$, whose integration over each time interval determines the probability of a leak occurring.
    \item[\textbf{(iv)}] \textbf{Uncertain thrust.} Variability in the thruster system leads to differences between the actual thrust, $Th_a$, and the commanded thrust, $Th_c$. This variability is modeled by a probability distribution $p_{Th}(Th_a)$, which captures the likelihood of various thrust values in response to $Th_c$. By integrating this distribution over the time intervals during which thrust is applied, one quantifies the overall impact of thrust uncertainty on maneuver performance. This probability is then incorporated into the state transition model to manage the effects of uncertain thrust.
\end{enumerate}

For the GRACE-FO mission, only thrust uncertainty is considered. Fuel leaks predominantly result from the attitude control system, which operates daily, so orbit-raising maneuvers have little effect on them. In addition, the spacecraft employs its attitude control system for all maneuvers, and no extra fuel is reserved for station keeping or collision avoidance, as these events were not part of the original design.



\subsection{Problem Formulation}

We now formalize the mission planning problem as an optimization task over a finite planning horizon. The objective is to maximize the total expected reward while ensuring that a critical safety specification is satisfied with high probability. We define a safety requirement expressed as a temporal logic formula that captures two key constraints:
\begin{enumerate}
    \item The spacecraft must not perform maneuvers more frequently than every three months.
    \item The spacecraft must never drop below an altitude of 300 km.
\end{enumerate}

The optimization problem is formulated as follows:
\begin{subequations}
\begin{align}
    \max_{\pi} & \qquad \mathbb{E} \left[ \sum_{h=1}^{H} R(s,a,h) \right] \label{eq:specification_value_function} \\
    \text{s.t.} & \qquad  \mathrm{Pr}^{\pi}( s_0 \models \varphi) \geq 1 - \delta, \label{eq:main_prob_2}
\end{align}
\end{subequations}
where$ R(s,a,h)$ is the reward obtained at time step $h$, $s_0$ is the initial state, $\varphi$ is the temporal logic safety specification, and $\delta$ denotes the maximum allowable probability of a safety violation.
The informal safety specification can be read as follows:
\begin{quote}
\textit{``The spacecraft does not perform a maneuver more often than every 3 months, and it never drops below 300 km."}
\end{quote}

To encode this requirement formally in a temporal logic specification, we define the following atomic propositions:
\begin{itemize}
    \item \(\texttt{maneuver}\): Evaluates to \texttt{true} when the spacecraft performs a maneuver at the current time step.
    \item \(\texttt{altitude300}\): Evaluates to \texttt{true} when the spacecraft's altitude is below 300\,km.
\end{itemize}

A possible temporal logic formula is:
\begin{align}
\begin{split}
\Box \Bigl[ \bigl(\texttt{maneuver} \rightarrow \neg (X\,\texttt{maneuver} \vee XX\,\texttt{maneuver})\bigr) \land
\neg \texttt{altitude300} \Bigr],
\end{split}\label{temporal_spec}
\end{align}
which reads as: “It is always the case that if a maneuver is executed, no maneuver is performed in the next two time steps, and the spacecraft’s altitude remains above 300 km.”

This formulation guarantees that maneuvers are sufficiently spaced in time and that the spacecraft operates within the safe altitude range throughout the mission.

\section{Experimental Results}

The proposed planning method is applied to the GRACE-FO mission, which we assume is scheduled to operate from May 2018 until December 2029. This operating period provides an extra year beyond the planned launch of GRACE-C in December 2028 to accommodate possible launch delays or initial measurement issues. At the start of the mission, the spacecraft is positioned at an altitude of 490\,km and is allocated 5\,kg of fuel for orbit-raising maneuvers.

The discretizations form the basis of the MDP model, in which each state is defined by the discretized altitude, fuel level, and time since the last maneuver. The available actions correspond to the feasible altitude changes. Accurate capture of small changes in altitude is achieved by dividing the altitude range between 300\,km and 500\,km into 1500 intervals. The fuel level is discretized into 500 states to model fuel consumption precisely. An additional state variable tracks the time elapsed since the last maneuver; this enforces the safety requirement for sufficient spacing between maneuvers. An action can be taken every month. To accommodate the low temporal resolution, the available maneuvers are chosen using a geometric progression, which permits various magnitudes of altitude increases.

\subsection{Deterministic Planning}
\begin{figure}
    \centering
    \subfigure[Deterministic altitude evolution]{%
         \includegraphics[width=0.495\textwidth]{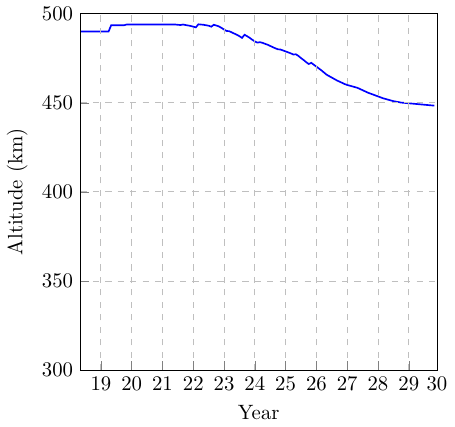}%
         \label{fig:altitude_deterministic}%
    }
    \hfill
    \subfigure[Deterministic fuel evolution]{%
         \includegraphics[width=0.47\textwidth]{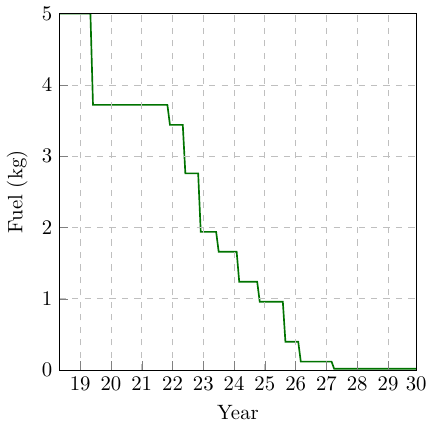}%
         \label{fig:fuel_deterministic}%
    }
    \caption{Deterministic altitude and fuel evolution (bottom) when solving the MDP with deterministic transitions.}
    \label{fig:mdp_deterministic}
\end{figure}

In the first set of simulations, the solar flux is assumed to be fixed at its expected value (as provided by NOAA\cite{miesch2024solar}), and the cost for performing an orbit-raising maneuver is treated as deterministic. With these assumptions, the MDP is solved using a backward-recursion scheme. Figure~\ref{fig:mdp_deterministic} illustrates the deterministic evolution of altitude and fuel consumption. In this simulation, the spacecraft's altitude increases in steps, reaching a final altitude slightly below 450\,km, while maneuvers occur approximately every 6 months. These results indicate that the solution consistently maintains the spacecraft above 300\,km.

In contrast, traditional mission planning methods typically schedule maneuvers at fixed intervals without considering the current state of the spacecraft. Such classic approaches spread maneuvers evenly over time, which can lead to unnecessary fuel consumption or insufficient altitude control. The MDP-based method, however, adapts the timing and magnitude of maneuvers based on the current altitude and fuel level. This dynamic scheduling helps to optimize fuel usage while ensuring that safety requirements are met more consistently.

\subsection{Planning Under Uncertainties}
\begin{figure}
    \includegraphics[width=\linewidth]{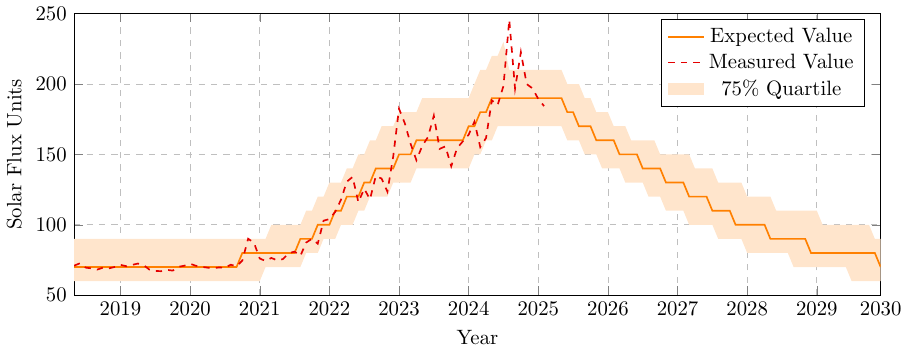}
    \caption{Solar flux evolution}
    \label{fig:solar_flux}
    \subfigure[Altitude evolution with uncertainties]{%
        \includegraphics[width=0.475\textwidth]{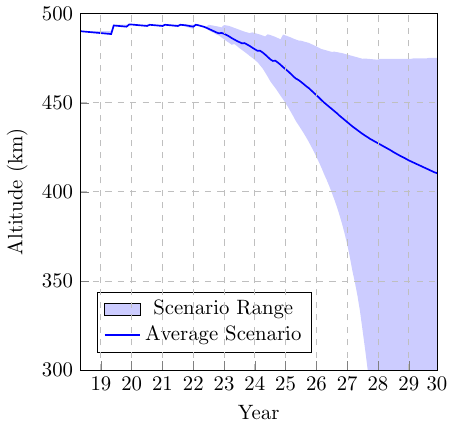}%
.    }
    \hfill
    \subfigure[Fuel evolution with uncertainties]{%
        \includegraphics[width=0.45\textwidth]{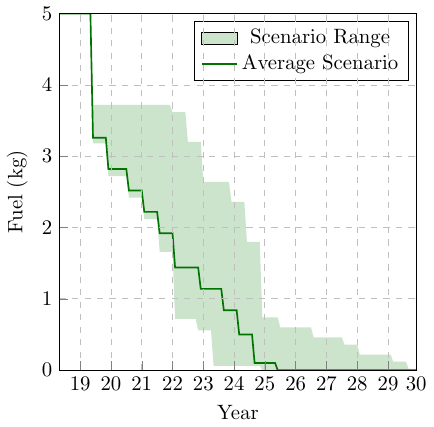}%
    }
    \caption{Altitude and fuel evolution when solving the MDP with uncertain transitions}
    \label{fig:mdp_solution}
        \centering
        \subfigure[Probability distribution for the final altitude]{%
        \includegraphics[width=0.47\textwidth]{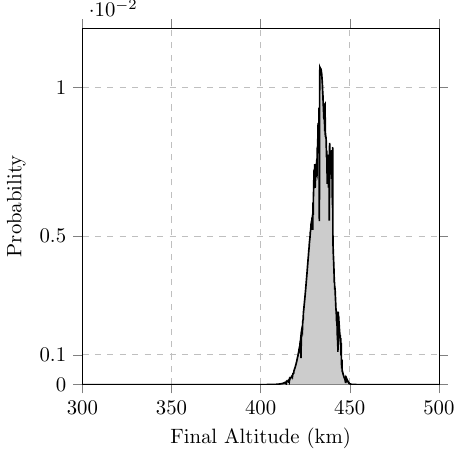}%
    }
    \hfill
    \subfigure[Cumulative distribution for the final altitude]{%
        \includegraphics[width=0.47\textwidth]{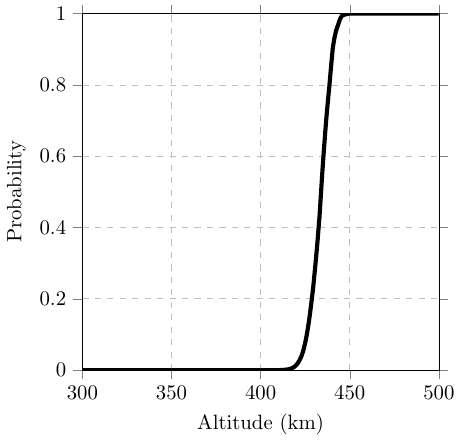}%
    }
    \caption{Probabilistic verification for the final altitude following the policy $\pi$ using the initial condition of GRACE-FO at its launch}
    \label{fig:verification}
\end{figure}

Next, we incorporate uncertainties into the model by treating both the solar flux and the outcomes of the orbit-raising maneuvers as random variables. Figure~\ref{fig:solar_flux} shows the evolution of the solar flux over time, with the 75\% quartile range highlighted in orange; these discretized distributions are integrated into the model. We then evaluate state transitions by considering best-case, worst-case, and average scenarios based on variations in solar flux and orbit-raising cost. Figure~\ref{fig:mdp_solution} displays a range of possible trajectories for altitude and fuel consumption. As expected, the worst-case scenario leads to fuel depletion early in 2025, while the best-case scenario extends fuel availability until 2029. The final altitude varies widely, from below 300\,km up to 475\,km, with an average of 410\,km. Although the average behavior appears to satisfy the temporal specification, this plot alone does not provide full probabilistic safety guarantees. The results suggest possible violations of the requirement to maintain an altitude above 300 km, motivating us to turn next to safety verification.

\subsection{Safety Verification}
We study the satisfaction of the safety requirement by computing the value function associated with the temporal logic specification. It ensures that the altitude never falls below 300\,km and that maneuvers are not executed too frequently. Figure~\ref{fig:verification} shows the probability distribution for the final altitude and its cumulative density function. The results demonstrate that the probability of falling below 300\,km is nearly zero, with a 95\% chance that the final altitude will exceed 425\,km. Analysis of the optimal policy further confirms that if a maneuver occurs within a recent period (specifically, within the last 6 months), no additional orbit-raising maneuver is scheduled.

These results underscore the strength of the MDP framework in enforcing strict safety constraints while optimizing performance. Unlike traditional planning methods that do not explicitly verify safety through continuous feedback, the value function in this approach provides a clear measure of safety compliance. The near-zero risk of dropping below the minimum altitude and the high probability of maintaining an altitude above 425\,km indicate that the policy not only meets but exceeds safety requirements. 
\begin{figure}
    \subfigure[Altitude evolution with uncertainties]{%
        \includegraphics[width=0.49\textwidth]{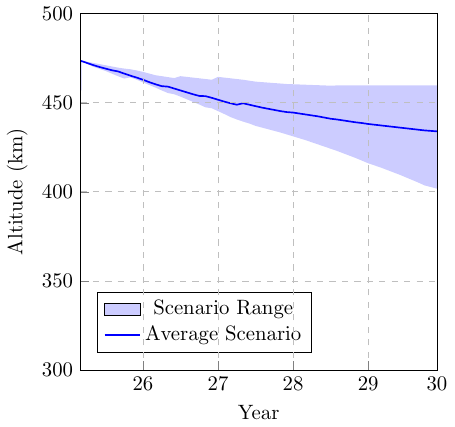}%
    }
    \hfill
    \subfigure[Fuel evolution with uncertainties]{%
        \includegraphics[width=0.49\textwidth]{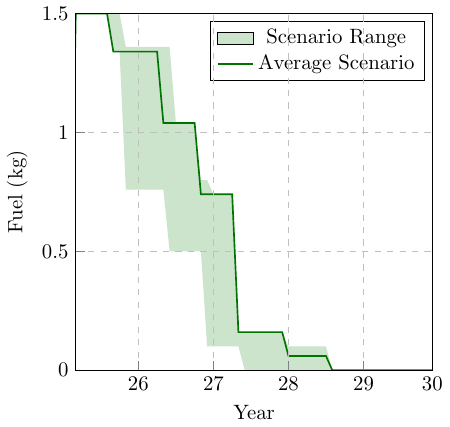}%
    }
    \caption{Altitude and fuel evolution when solving the MDP with uncertain transitions using GRACE-FO state in March 2025 as the initial condition}
    \label{fig:now}
\end{figure}
\subsection{Current-State Planning}
We can also use the previously synthesized policy to update the mission forecast by re-simulating  from the spacecraft’s current state, reducing uncertainty about future trajectories. For instance, in early 2025, with only 1.5 kg of propellant remaining, propagating the existing policy yields a maneuver sequence in Fig.~\ref{fig:now} that maintains altitude safely above 300 km, and in fact no scenarios fall below 400 km. This demonstrates how the method can incorporate real mission feedback to refine expectations about future performance under the existing policy.

In addition, we can resolve the MDP when improved solar-activity forecasts become available. By updating the transition probabilities, this would enable the resulting policies to adapt to evolving environmental predictions, reducing conservatism compared to fixed-scenario approaches and improving robustness under uncertainty.

\section{Conclusion}
We presented a finite-horizon MDP framework for probabilistic mission planning and safety verification, applied to the GRACE-FO mission. By embedding environmental and system uncertainties directly into the state-transition model, we get a maneuver strategy that balances fuel efficiency, data quality, and safety under uncertainty.

While the results demonstrate the feasibility of MDP-based planning on a simplified orbital model, an important next step is to test robustness against the modeling gap between this abstraction and the full mission dynamics. This extension requires coupling the MDP policies with higher-fidelity orbital simulators and assimilating realistic thruster performance data. Such integration will allow systematic evaluation of how well the synthesized policies transfer from the reduced-order model to operationally relevant scenarios, and whether safety guarantees remain valid under unmodeled effects. Taken together, these results point toward reliable autonomous decision-making for spacecraft, ensuring that guarantees derived from simplified models remain valid in real space operations.

\section{Aknowledgment}
This work is supported by NASA Grant 80NSSC23K1343.

\bibliographystyle{AAS_publication}   
\bibliography{references}   

\end{document}